\documentclass[amsmath,aps,showpacs,a4paper,10pt]{revtex4}

 \usepackage{epsf}
 \usepackage{graphicx}    

\begin{document}

 \newcommand{\be}[1]{\begin{equation}\label{#1}}
 \newcommand{\ee}{\end{equation}}
 \newcommand{\bea}{\begin{eqnarray}}
 \newcommand{\eea}{\end{eqnarray}}
 \def\disp{\displaystyle}

 \def\gsim{ \lower .75ex \hbox{$\sim$} \llap{\raise .27ex \hbox{$>$}} }
 \def\lsim{ \lower .75ex \hbox{$\sim$} \llap{\raise .27ex \hbox{$<$}} }

 \begin{titlepage}

 \begin{flushright}
 arXiv:0902.2030
 \end{flushright}

 \title{\Large \bf Modified Holographic Dark Energy}

 \author{Hao~Wei\,}
 \email[\,email address:\ ]{haowei@bit.edu.cn}
 \affiliation{Department of Physics, Beijing Institute
 of Technology, Beijing 100081, China}

 \begin{abstract}\vspace{1cm}
 \centerline{\bf ABSTRACT}\vspace{2mm}
In this work, motivated by the energy bound suggested
 by Cohen {\it et al.}, we propose the modified holographic
 dark energy (MHDE) model. Choosing the IR cut-off
 $L=R_{\rm CC}$ and considering the parameterizations
 $n^2=2-\lambda a$, $n^2=2-3\lambda a^2/\left(1+3a^2\right)$
 and $n^2=2-\lambda a^2/\left(\beta+a^2\right)$, we derive all
 the physical quantities of the non-saturated MHDE model
 analytically. We find that the non-saturated MHDE models
 with the parameterizations $n^2=2-\lambda a$ and
 $n^2=2-3\lambda a^2/\left(1+3a^2\right)$ are single-parameter
 models in practice. Also, we consider the cosmological
 constraints on the non-saturated MHDE, and find that it is
 well consistent with the observational data.
 \end{abstract}


 \maketitle

 \end{titlepage}

 \renewcommand{\baselinestretch}{1.6}


\section{Introduction}\label{sec1}
Recently, the holographic dark energy (HDE) has been considered
 as an interesting candidate of dark energy, which has been
 studied extensively in the literature. It was proposed from
 the well-known holographic principle~\cite{r1,r2} in the
 string theory. For a quantum gravity system, the local
 quantum field cannot contain too many degrees of freedom,
 otherwise the formation of black hole is inevitable and then
 the quantum field theory breaks down. In the black hole
 thermodynamics~\cite{r3,r4}, there is a maximum entropy in a
 box of size $L$, namely the Bekenstein-Hawking entropy bound
 $S_{BH}$, which scales as the area of the box $\sim L^2$,
 rather than the volume $\sim L^3$. To avoid the breakdown of
 the local quantum field theory, Cohen {\it et al.}~\cite{r5}
 proposed a more restrictive bound, namely the energy bound.
 If $\rho_\Lambda$ is the quantum zero-point energy density
 caused by a short distance cut-off, the total energy in a box
 of size $L$ cannot exceed the mass of a black hole of the same
 size~\cite{r5}, namely $\rho_\Lambda L^3\,\lsim\, m_p^2L$,
 where $m_p\equiv (8\pi G)^{-1/2}$ is the reduced Planck mass.
 The largest IR cut-off $L$ is the one saturating the
 inequality. Thus,
 \be{eq1}
 \rho_\Lambda=3n^2 m_p^2 L^{-2},
 \ee
 where the numerical constant $3n^2$ is introduced for
 convenience. If we choose $L$ as the size of the universe,
 for instance the Hubble horizon $H^{-1}$, the resulting
 $\rho_\Lambda$ is comparable to the observational density
 of dark energy~\cite{r6,r7}. However, Hsu~\cite{r7} pointed
 out that in this case the resulting equation-of-state
 parameter~(EoS) is equal to zero, which cannot accelerate the
 expansion of the universe. The other possibility~\cite{r8}
 is to choose $L$ as the particle horizon
 \be{eq2}
 R_H\equiv a\int_0^t\frac{d\tilde{t}}{a}
 =a\int_0^a\frac{d\tilde{a}}{H\tilde{a}^2}\,,
 \ee
 where $H\equiv\dot{a}/a$ is the Hubble parameter;
 $a=(1+z)^{-1}$ is the scale factor of the universe; $z$ is the
 redshift; we have set $a_0=1$; the subscript ``0'' indicates
 the present value of the corresponding quantity; a dot denotes
 the derivative with respect to cosmic time $t$. However, it is
 easy to find that the EoS is always larger than $-1/3$ and
 also cannot accelerate the expansion of the universe in this
 case with $L=R_H$~\cite{r9}. To get an accelerating universe,
 Li proposed in~\cite{r9} to choose $L$ as the future event
 horizon
 \be{eq3}
 R_h\equiv a\int_t^\infty\frac{d\tilde{t}}{a}
 =a\int_a^\infty\frac{d\tilde{a}}{H\tilde{a}^2}\,.
 \ee
 In this case, the EoS of the holographic dark energy can be
 less than $-1/3$~\cite{r9}. In fact, the version of HDE with
 $L=R_h$ is the one considered extensively in the literature.
 For a comprehensive list of references concerning HDE, we
 refer to e.g.~\cite{r10,r11,r12,r36,r38} and references
 therein.

In the literature, many authors extensively considered the
 HDE whose energy density was taken to be the one given in
 Eq.~(\ref{eq1}). Here, we instead propose to consider a
 modified version of HDE. Notice that the energy bound proposed
 in~\cite{r5} requires that the {\em total} energy in a box of
 size $L$ cannot exceed the mass of a black hole of the same
 size. Therefore, it is reasonable to argue that when HDE
 coexists with matter, the total energy should satisfy
 the energy bound, namely, $\rho_{tot} L^3\,\lsim\, m_p^2L$.
 Correspondingly, the largest IR cut-off $L$ is the one
 saturating the inequality. Thus,
 \be{eq4}
  \rho_{tot}=3n^2 m_p^2 L^{-2},
 \ee
 where $\rho_{tot}=\rho_m+\rho_\Lambda$ is the total energy
 density; $\rho_m=\rho_{m0} a^{-3}$ is the energy density of matter.
 Therefore, the energy density of the modified HDE is given by
 $\rho_\Lambda=\rho_{tot}-\rho_m=3n^2 m_p^2 L^{-2}-\rho_{m0} a^{-3}$
 instead. Note that we do not consider the radiation in this
 work, since almost all the cosmological observations probe the
 cosmic history after the matter-radiation equality. Using the
 Friedmann equation $3m_p^2 H^2=\rho_{tot}$ and
 Eq.~(\ref{eq4}), it is easy to find that $H^2 L^2=n^2$. Note
 that we consider a flat universe throughout. Since $n$ is
 given in the form of $n^2$ throughout, we only consider the
 positive $n$ in this work. So, we have
 \be{eq5}
 HL=n.
 \ee
 This very simple equation is our starting point for the
 modified HDE (hereafter MHDE). Naturally, the next step is
 to choose an appropriate $L$.


\section{Choosing an appropriate IR cut-off}\label{sec2}
The most natural choice for $L$ is $L=H^{-1}$. In this case,
 one should have $n=1$ from Eq.~(\ref{eq5}). So, the
 requirement of the holographic principle, namely
 Eq.~(\ref{eq4}), is nothing but the familiar Friedmann
 equation. There is no new input. Therefore, the choice
 $L=H^{-1}$ is trivial.

The second choice for $L$ is the particle horizon $R_H$ in
 Eq.~(\ref{eq2}). For $L=R_H$, we have
 \be{eq6}
 \frac{dL}{da}=\frac{L}{a}+\frac{1}{Ha}=
 \frac{L}{a}\left(1+\frac{1}{n}\right),
 \ee
 which can be regarded as a differential equation for $L$ with
 respect to $a$. From Eq.~(\ref{eq6}), one can find that
 $L\propto a^{1+n^{-1}}$ and $H=n/L\propto a^{-1-n^{-1}}$.
 Noting that in the matter-dominated epoch $H\propto a^{-3/2}$,
 MHDE with $L=R_H$ can describe the matter-dominated epoch
 when $n=2$. This is a good news. However, we should consider
 the second thing. As is well known,
 \be{eq7}
 \frac{\ddot{a}}{a}=\dot{H}+H^2=H^2+Ha\frac{dH}{da}\,.
 \ee
 Substituting $H\propto a^{-1-n^{-1}}$ into Eq.~(\ref{eq7}),
 we have $\ddot{a}/a=-n^{-1}H^2<0$. Therefore, the universe
 cannot be accelerated, and hence the choice $L=R_H$ is
 unsuitable.

The third choice for $L$ is the future event horizon $R_h$ in
 Eq.~(\ref{eq3}). For $L=R_h$, we have
 \be{eq8}
 \frac{dL}{da}=\frac{L}{a}-\frac{1}{Ha}=
 \frac{L}{a}\left(1-\frac{1}{n}\right).
 \ee
 So, it is easy to see that $L\propto a^{1-n^{-1}}$ and
 $H=n/L\propto a^{-1+n^{-1}}$. Unfortunately, it cannot
 describe the matter-dominated epoch in which
 $H\propto a^{-3/2}$. On the other hand, substituting
 $H\propto a^{-1+n^{-1}}$ into Eq.~(\ref{eq7}), we have
 $\ddot{a}/a=n^{-1}H^2>0$. Thus, the universe is accelerating,
 but cannot undergo a decelerated epoch. So, the choice
 $L=R_h$ is also unsuitable.

There is another choice. In~\cite{r13}, Gao {\it et al.}
 suggested that $L$ can be proportional to the Ricci scalar
 curvature radius. The resulting HDE was called Ricci dark
 energy in the literature. See e.g.~\cite{r14,r15,r16,r17}
 for works on the so-called Ricci dark energy.
 In~\cite{r13,r14,r15,r16,r17}, there is no physical motivation
 to this proposal for $L$ in fact. Recently, in~\cite{r18},
 Cai {\it et al.} found that the Jeans length $R_{\rm CC}$
 which is determined by $R_{\rm CC}^{-2}=\dot{H}+2H^2$ gives
 the causal connection scale of perturbations in the flat
 universe. Since the Ricci scalar is also proportional to
 $\dot{H}+2H^2$ in the flat universe, Cai {\it et al.} in fact
 found the physical motivation for the Ricci dark energy. Here,
 we take $L=R_{\rm CC}$ as the fourth choice for MHDE. In
 this case, from Eq.~(\ref{eq5}), we obtain
 \be{eq9}
 H^2=n^2L^{-2}=n^2\left(\dot{H}+2H^2\right).
 \ee
 It can be recast as a differential equation for $H$ with
 respect to $a$, namely
 \be{eq10}
 \frac{dH}{da}=\left(\frac{1}{n^2}-2\right)\frac{H}{a}\,.
 \ee
 So, we find that $H\propto a^{n^{-2}-2}$. When $n^2=2$, MHDE
 with $L=R_{\rm CC}$ can describe the matter-dominated epoch
 in which $H\propto a^{-3/2}$. On the other hand, substituting
 $H\propto a^{n^{-2}-2}$ into Eq.~(\ref{eq7}), we find that
 \be{eq11}
 \frac{\ddot{a}}{a}=\left(\frac{1}{n^2}-1\right)H^2.
 \ee
 Thus, the universe is decelerating when $n>1$, whereas the
 universe is accelerating when $n<1$. So, the choice
 $L=R_{\rm CC}$ gives the possibility for a cosmologically
 feasible MHDE model.

However, in the saturated MHDE model, $n$ is constant. For a
 given $n$, it is impossible to describe both the decelerated
 and accelerated phases, even in the case with $L=R_{\rm CC}$.
 To build a cosmologically feasible MHDE model, we should go
 further.


\section{Non-saturated MHDE}\label{sec3}
In the derivation of HDE mentioned in Sec.~\ref{sec1}, the
 holographic bound was chosen to be saturated. However, this
 is not necessary. In~\cite{r19}, Guberina {\it et al.}
 proposed the so-called non-saturated HDE, in which the
 parameter $n=n(t)$ is a function of cosmic time $t$, rather
 than a constant. Similarly, here we consider the non-saturated
 MHDE with $L=R_{\rm CC}$, in which $n=n(a)$ is a function of
 scale factor $a$. To solve Eq.~(\ref{eq10}), we should specify
 the explicit form of $n(a)$. Similar to the well-known EoS
 parameterization $w=w_0+w_a(1-a)=w_1-w_2\,a$ considered
 extensively in the literature~\cite{r20}, we parameterize
 $n^2=\alpha-\lambda a$, where $\alpha$ and $\lambda$ are
 constants. In fact, one can regard this type of
 parameterization as a linearized expansion with respect to
 scale factor $a$. As shown in Sec.~\ref{sec2}, when
 $n^2=2$, we find that MHDE with $L=R_{\rm CC}$ can describe
 the matter-dominated epoch in which $H\propto a^{-3/2}$.
 So, we get $\alpha=2$ by requiring $n^2\to 2$ when $a\to 0$.
 From now on, we consider the non-saturated MHDE in which
 $L=R_{\rm CC}$ and
 \be{eq12}
 n^2=2-\lambda a.
 \ee
 Noting that $n^2>0$, we should require $\lambda<2$ at least
 to describe the cosmic history up to now, namely
 $0\le z<\infty$. Substituting Eq.~(\ref{eq12}) into
 Eq.~(\ref{eq10}), we can recast it as
 \be{eq13}
 \frac{dH}{da}=\left(\frac{1}{2-\lambda a}-2\right)\frac{H}{a}\,.
 \ee
 The solution reads
 \be{eq14}
 H=H_0\left(2-\lambda\right)^{1/2} a^{-3/2}
 \left(2-\lambda a\right)^{-1/2},
 \ee
 where $H_0=H(z=0)$ is the Hubble constant. Obviously,
 $H\propto a^{-3/2}$ when $a\to 0$, and hence this model can
 describe the matter-dominated epoch. For convenience, we
 rewrite Eq.~(\ref{eq14}) to
 \be{eq15}
 E^2=\left(2-\lambda\right)\left(1+z\right)^3\left(2-
 \frac{\lambda}{1+z}\right)^{-1},
 \ee
 where $E\equiv H/H_0$. Substituting Eq.~(\ref{eq14}) into
 Eq.~(\ref{eq7}), we can obtain the deceleration parameter
 \be{eq16}
 q\equiv -\frac{\ddot{a}}{aH^2}=
 \frac{1}{2}\left(1-\frac{\lambda a}{2-\lambda a}\right).
 \ee
 Obviously, the universe is decelerating when $\lambda a<1$,
 whereas the universe is accelerating when $\lambda a>1$.
 The transition occurs when $\lambda a=1$. Therefore, the
 transition redshift is given by
 \be{eq17}
 z_t=\lambda-1.
 \ee
 From the Friedmann equation
 $3m_p^2H^2=\rho_{tot}=\rho_m+\rho_\Lambda$, we have
 $\rho_\Lambda=3m_p^2H^2-\rho_{m0}a^{-3}$, or equivalently
 \be{eq18}
 \tilde{\rho}_\Lambda\equiv\frac{\rho_\Lambda}{3m_p^2H_0^2}=
 \left(2-\lambda\right)\left(1+z\right)^3\left(2-
 \frac{\lambda}{1+z}\right)^{-1}-\Omega_{m0}\left(1+z\right)^3.
 \ee
 The fractional energy density
 $\Omega_i\equiv\rho_i/\left(3m_p^2H^2\right)$, where $i=m$
 and $\Lambda$. In particular, we find that
 \be{eq19}
 \Omega_m=\frac{\Omega_{m0}a^{-3}}{E^2}= \frac{\Omega_{m0}}{2-
 \lambda}\left(2-\frac{\lambda}{1+z}\right).
 \ee
 Requiring $\Omega_m\to 1.0$ (namely matter dominated) when
 $z\to\infty$, we have
 \be{eq20}
 \Omega_{m0}=1-\frac{\lambda}{2}\,.
 \ee
 Unlike the other versions of HDE considered in the literature,
 $\Omega_{m0}$ is {\em not} an independent parameter in the
 non-saturated MHDE model. This is an important result. Using
 Eq.~(\ref{eq20}), we finally obtain
 \bea
 &&\Omega_m=1-\frac{\lambda}{2\left(1+z\right)}\,,\label{eq21}\\
 &&\Omega_\Lambda=1-\Omega_m=
 \frac{\lambda}{2\left(1+z\right)}\,.\label{eq22}
 \eea
 \newpage 
 \noindent Interestingly, we find that $\Omega_\Lambda=\Omega_m$ at
 the transition redshift $z_t$ given in Eq.~(\ref{eq17}) which
 is determined by $q=0$. Provided that $\lambda$ is of order
 unity, the cosmological coincidence problem could be
 alleviated. On the other hand, it is easy to find the
 effective EoS as (see e.g.~\cite{r21,r37})
 \be{eq23}
 w_{\rm eff}\equiv\frac{p_{tot}}{\rho_{tot}}
 =-1-\frac{2}{3}\frac{\dot{H}}{H^2}
 =\frac{1}{3}\left(2q-1\right)
 =-\frac{\lambda a}{\,3\left(2-\lambda a\right)}\,,
 \ee
 Noting that
 $w_{\rm eff}=\Omega_m w_m+\Omega_\Lambda w_\Lambda$ and
 $w_m=0$, the EoS of HDE reads
 \be{eq24}
 w_\Lambda=\frac{w_{\rm eff}}{\Omega_\Lambda}\,,
 \ee
 where $\Omega_\Lambda$ is given in Eq.~(\ref{eq22}). If
 $\lambda$ is given, all the physical quantities can be derived
 accordingly, thanks to the important result Eq.~(\ref{eq20}).
 In fact, to our knowledge, the non-saturated MHDE model
 considered in this work is the fourth {\em single-parameter}
 cosmological model besides the well-known flat $\Lambda$CDM
 model, the flat DGP braneworld model~\cite{r22} and the new
 agegraphic dark energy (NADE) model~\cite{r23}.


\section{Cosmological constraints on the non-saturated MHDE}\label{sec4}
Here, we consider the cosmological constraints on the
 non-saturated MHDE. In~\cite{r24,r25}, Supernova
 Cosmology Project (SCP) collaboration released their new
 dataset of type~Ia supernovae~(SNIa), which was called
 Union compilation. The Union compilation contains 414 SNIa
 and reduces to 307 SNIa after selection cuts~\cite{r24}.

We perform a $\chi^2$ analysis to obtain the constraints on the
 single parameter $\lambda$ of the non-saturated MHDE model.
 The data points of the 307 Union SNIa compiled in~\cite{r24}
 are given in terms of the distance modulus $\mu_{obs}(z_i)$.
 On the other hand, the theoretical distance modulus is
 defined as
 \be{eq25}
 \mu_{th}(z_i)\equiv 5\log_{10}D_L(z_i)+\mu_0\,,
 \ee
 where $\mu_0\equiv 42.38-5\log_{10}h$ and $h$ is the Hubble
 constant $H_0$ in units of $100~{\rm km/s/Mpc}$, whereas
 \be{eq26}
 D_L(z)=(1+z)\int_0^z \frac{d\tilde{z}}{E(\tilde{z};\lambda)}\,,
 \ee
 The $\chi^2$ from the 307 Union SNIa are given by
 \be{eq27}
 \chi^2_{\mu}(\lambda)=\sum\limits_{i}\frac{\left[
 \mu_{obs}(z_i)-\mu_{th}(z_i)\right]^2}{\sigma^2(z_i)}\,,
 \ee
 where $\sigma$ is the corresponding $1\sigma$ error. The parameter
 $\mu_0$ is a nuisance parameter but it is independent of the data
 points. One can perform an uniform marginalization over $\mu_0$.
 However, there is an alternative way. Following~\cite{r26,r27}, the
 minimization with respect to $\mu_0$ can be made by expanding the
 $\chi^2_{\mu}$ of Eq.~(\ref{eq27}) with respect to $\mu_0$ as
 \be{eq28}
 \chi^2_{\mu}(\lambda)=\tilde{A}-2\mu_0\tilde{B}+\mu_0^2\tilde{C}\,,
 \ee
 where
 $$\tilde{A}(\lambda)=\sum\limits_{i}\frac{\left[
 \mu_{obs}(z_i)-\mu_{th}(z_i;\mu_0=0,\lambda)
 \right]^2}{\sigma_{\mu_{obs}}^2(z_i)}\,,$$
 $$\tilde{B}(\lambda)=\sum\limits_{i}\frac{\mu_{obs}(z_i)
 -\mu_{th}(z_i;\mu_0=0,\lambda)}{\sigma_{\mu_{obs}}^2(z_i)}\,,
 ~~~~~~~~~~
 \tilde{C}=\sum\limits_{i}\frac{1}{\sigma_{\mu_{obs}}^2(z_i)}\,.$$
 Eq.~(\ref{eq28}) has a minimum for
 $\mu_0=\tilde{B}/\tilde{C}$ at
 \be{eq29}
 \tilde{\chi}^2_{\mu}(\lambda)=
 \tilde{A}(\lambda)-\frac{\tilde{B}(\lambda)^2}{\tilde{C}}\,.
 \ee
 Since $\chi^2_{\mu,\,min}=\tilde{\chi}^2_{\mu,\,min}$
 obviously, we can instead minimize $\tilde{\chi}^2_{\mu}$
 which is independent of $\mu_0$. Note that the above
 summations are over the 307 Union SNIa compiled in~\cite{r24}.
 It is worth noting that the corresponding $h$ can be determined by
 $\mu_0=\tilde{B}/\tilde{C}$ for the best-fit parameter.

\begin{center}
 \begin{figure}[tbhp]
 \centering
 \includegraphics[width=0.846\textwidth]{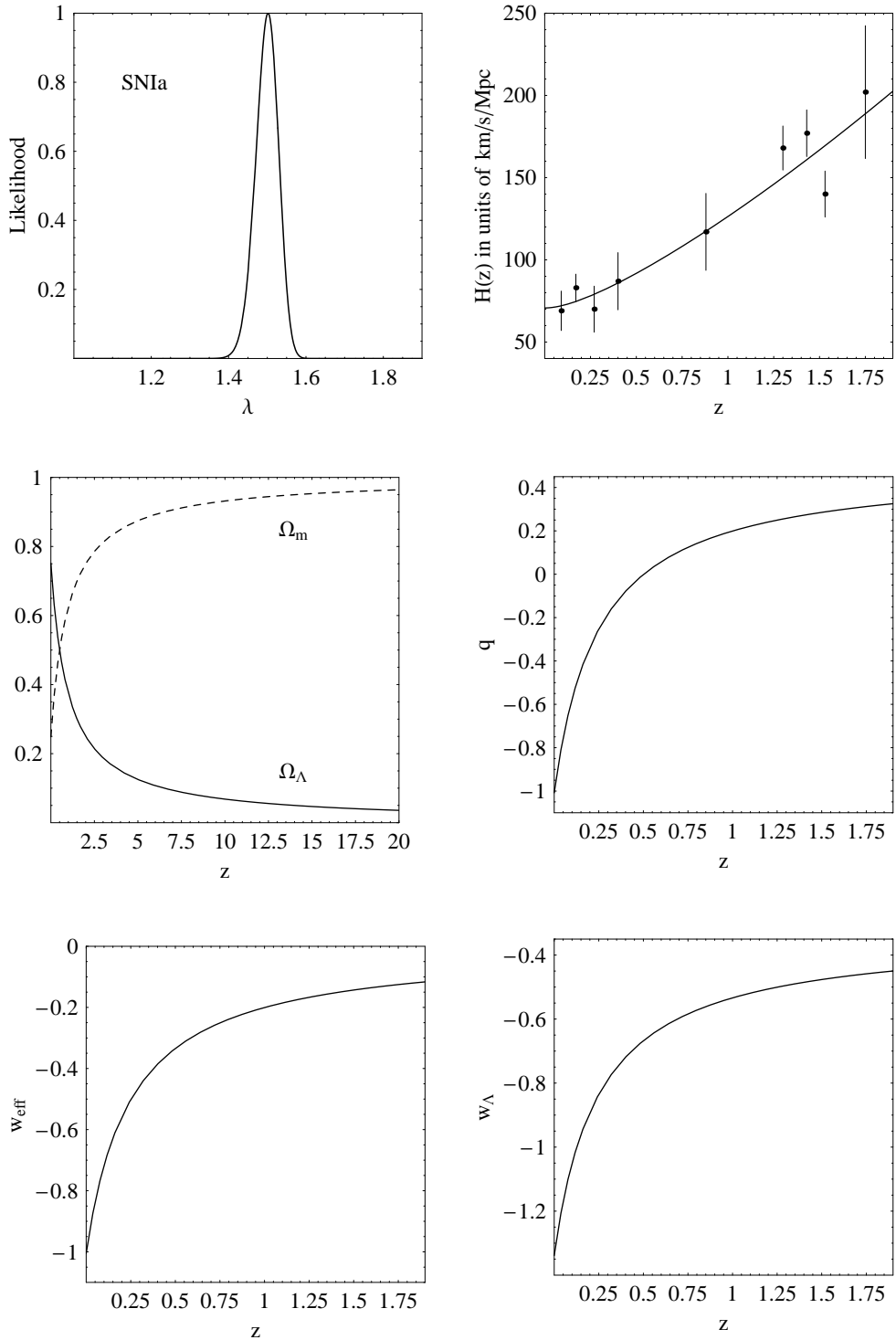}
 \caption{\label{fig1}
 The top-left panel is the likelihood ${\cal L}\propto e^{-\chi^2/2}$
 versus $\lambda$. The other panels are $H$, $\Omega_m$,
 $\Omega_\Lambda$, $q$, $w_{\rm eff}$ and $w_\Lambda$ as functions
 of redshift $z$ for the best-fit $\lambda$. These results are
 obtained by using the SNIa data only.}
 \end{figure}
 \end{center}

At first, we consider the cosmological constraints on the
 non-saturated MHDE, by using the 307 Union SNIa dataset only.
 In the top-left panel of Fig.~\ref{fig1}, we present the likelihood
 ${\cal L}\propto e^{-\chi^2/2}$ versus $\lambda$. We find
 \\ \vspace{-3mm}\\ 
 that the best-fit model parameter is
 $\lambda=1.502^{+0.028}_{-0.030}$ (with $1\sigma$ uncertainty)
 or $\lambda=1.502^{+0.054}_{-0.063}$ (with $2\sigma$ uncertainty),
 while $\chi^2_{min}=311.272$. The corresponding $h=0.707$. We
 plot $H(z)$ for the best-fit parameter $\lambda$ in the top-right panel
 of Fig.~\ref{fig1}. We also present the observational $H(z)$
 data from~\cite{r28} (see also \cite{r21,r29,r30}) for comparison.
 It is easy to see that $H(z)$ of the non-saturated MHDE is
 well consistent with the observational $H(z)$ data. In
 addition, $\Omega_m$, $\Omega_\Lambda$, $q$, $w_{\rm eff}$ and
 $w_\Lambda$ as functions of redshift $z$ for the best-fit
 $\lambda$ are also presented in Fig.~\ref{fig1}. Clearly,
 $\Omega_\Lambda$ and $\Omega_m$ are comparable since the near
 past, whereas $\Omega_{m0}=1-\lambda/2=0.249$ and
 $\Omega_{\Lambda 0}=\lambda/2=0.751$. We see that $q$ changed
 from $q>0$ to $q<0$ at $z_t=\lambda-1=0.502$. On the other
 hand, $w_\Lambda$ crossed the so-called phantom divide $w=-1$.

In the literature, the shift parameter $R$ from the cosmic microwave
 background (CMB) anisotropy, and the distance parameter $A$ of
 the measurement of the baryon acoustic oscillation (BAO) peak
 in the distribution of SDSS luminous red galaxies~\cite{r31},
 are also used extensively in obtaining the cosmological
 constraints. The shift parameter $R$ is defined
 by~\cite{r32,r33}
 \be{eq30}
 R\equiv\Omega_{m0}^{1/2}\int_0^{z_\ast}
 \frac{d\tilde{z}}{E(\tilde{z})}\,,
 \ee
 where the redshift of recombination $z_\ast=1090$ which has
 been updated in the 5-year data of Wilkinson Microwave
 Anisotropy Probe (WMAP5)~\cite{r34}. The shift parameter $R$
 relates the angular diameter distance to the last scattering
 surface, the comoving size of the sound horizon at $z_\ast$
 and the angular scale of the first acoustic peak in the CMB
 power spectrum of the temperature fluctuations~\cite{r32,r33}.
 The value of $R$ has been updated to $1.710\pm 0.019$ in
 WMAP5~\cite{r34}. The distance parameter $A$ is given by
 \be{eq31}
 A\equiv\Omega_{m0}^{1/2}E(z_b)^{-1/3}\left[\frac{1}{z_b}
 \int_0^{z_b}\frac{d\tilde{z}}{E(\tilde{z})}\right]^{2/3},
 \ee
 where $z_b=0.35$. In~\cite{r35}, the value of $A$ has been
 determined to be $0.469\,(n_s/0.98)^{-0.35}\pm 0.017$. Here
 the scalar spectral index $n_s$ is taken to be $0.960$ which
 has been updated in WMAP5~\cite{r34}.

\begin{center}
 \begin{figure}[tbhp]
 \centering
 \includegraphics[width=0.846\textwidth]{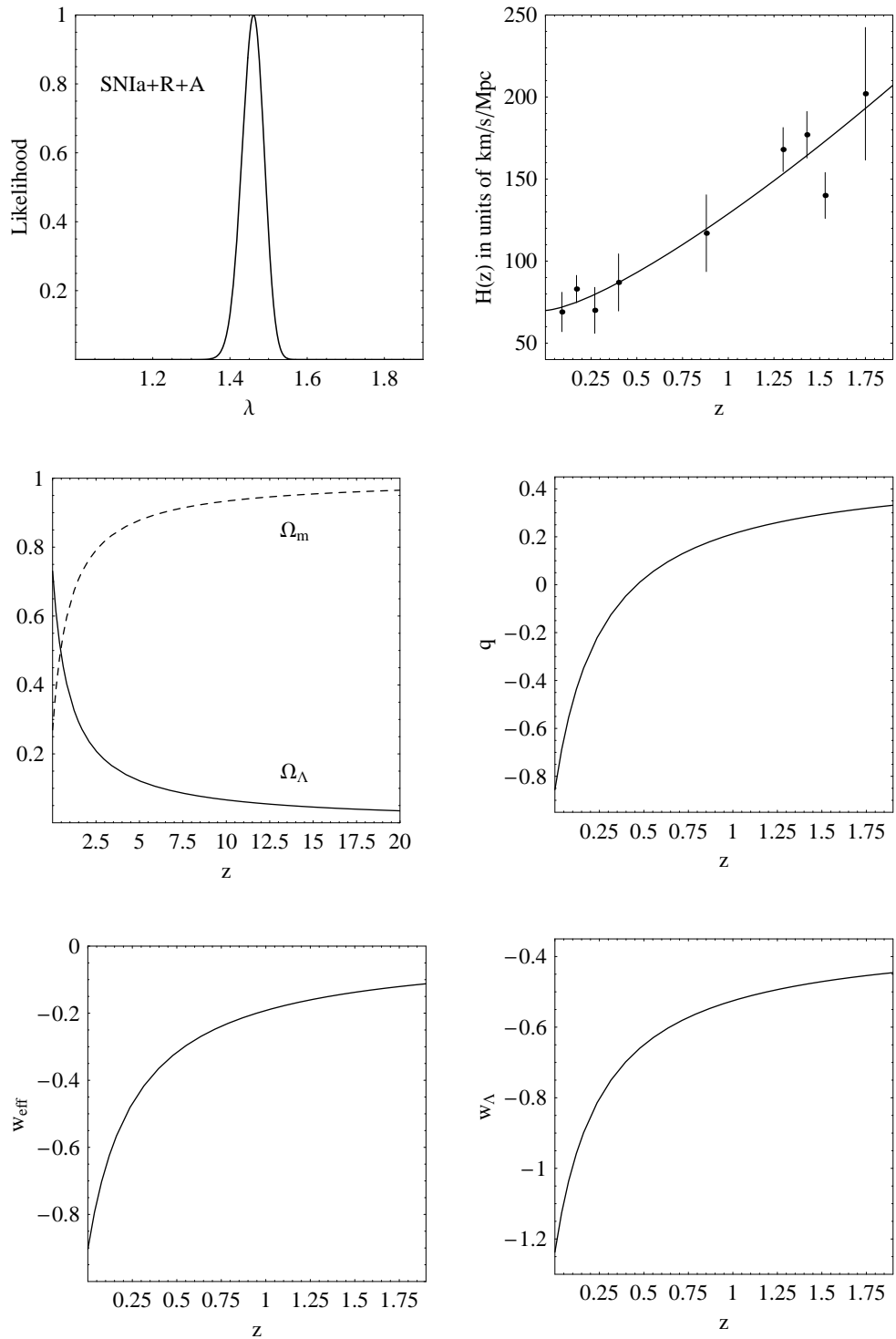}
 \caption{\label{fig2}
 The same as in Fig.~\ref{fig1}, except that these results are
 obtained by using the combined data of the 307 Union SNIa, the
 shift parameter $R$ of CMB and the distance parameter $A$ of BAO.}
 \end{figure}
 \end{center}

Now, we perform a joint $\chi^2$ analysis to obtain the cosmological
 constraints on the single parameter $\lambda$ of the non-saturated
 MHDE model, by using the combined data of the 307 Union SNIa,
 the shift parameter $R$ of CMB and the distance parameter $A$
 of BAO. Here, the total $\chi^2$ is given by
 \be{eq32}
 \chi^2=\tilde{\chi}^2_{\mu}+\chi^2_{CMB}+\chi^2_{BAO}\,,
 \ee
 where $\tilde{\chi}^2_{\mu}$ is given in Eq.~(\ref{eq29}),
 $\chi^2_{CMB}=(R-R_{obs})^2/\sigma_R^2$ and
 $\chi^2_{BAO}=(A-A_{obs})^2/\sigma_A^2$. The best-fit model
 parameter can be determined by minimizing the total $\chi^2$.
 In the top-left panel of Fig.~\ref{fig2}, we present the likelihood
 ${\cal L}\propto e^{-\chi^2/2}$ versus $\lambda$. We find that the
 best-fit model parameter is $\lambda=1.461^{+0.028}_{-0.029}$ (with
 $1\sigma$ uncertainty) or $\lambda=1.461^{+0.054}_{-0.061}$
 (with $2\sigma$ uncertainty), while $\chi^2_{min}=320.858$.
 The corresponding $h=0.699$. We also plot $H(z)$ for the best-fit
 $\lambda$ in the top-right panel of Fig.~\ref{fig2}. Again, it
 is easy to see that $H(z)$ of the non-saturated MHDE is well
 consistent with the observational $H(z)$ data. In addition,
 $\Omega_m$, $\Omega_\Lambda$, $q$, $w_{\rm eff}$ and $w_\Lambda$ as
 functions of redshift $z$ for the best-fit $\lambda$ are also
 presented in Fig.~\ref{fig2}. Clearly, $\Omega_\Lambda$ and
 $\Omega_m$ are comparable since the near past, whereas
 $\Omega_{m0}=1-\lambda/2=0.269$ and
 $\Omega_{\Lambda 0}=\lambda/2=0.731$. We see that $q$ changed
 from $q>0$ to $q<0$ at $z_t=\lambda-1=0.461$. On the other
 hand, $w_\Lambda$ crossed the so-called phantom divide $w=-1$.

As shown above, the non-saturated MHDE with the
 simplest parameterization $n^2=2-\lambda a$ is
 well consistent with the observational data in fact. This is
 fairly impressive, since all the physical quantities of this
 model are given by the simple and analytical expressions.


\vspace{-25mm}  


\section{Other parameterizations for the non-saturated MHDE}\label{sec5}
Although the simplest parameterization $n^2=2-\lambda a$ given
 in Eq.~(\ref{eq12}) has been shown that it is fairly
 successful to fit the cosmological observations, it is only
 valid for $a<2/\lambda$ which is required by $n^2>0$. So, we
 cannot use it to describe the future evolution of the
 universe. In fact, it is non-trivial to find a suitable
 parameterization of $n$ which is well-defined for the whole
 $0\leq a<\infty$. Naively, one can consider the simplest
 parameterization $n^2=2-\lambda a/\left(\beta+a\right)$.
 However, we find that it cannot be consistent with the
 combined observational data of the 307 Union SNIa, the shift
 parameter $R$ of CMB and the distance parameter $A$ of BAO.
 Here, we consider another parameterization
 $n^2=\alpha-\lambda a^2/\left(\beta+a^2\right)$,
 where $\alpha$, $\beta$ and $\lambda$ are constants. Again,
 we obtain $\alpha=2$ by requiring $n^2\to 2$ when $a\to 0$,
 in order to describe the matter-dominated epoch in which
 $H\propto a^{-3/2}$. From now on, we consider the
 non-saturated MHDE in which $L=R_{\rm CC}$ and
 \be{eq33}
 n^2=2-\frac{\lambda a^2}{\beta+a^2}\,.
 \ee
 We require $\beta\geq 0$ to avoid singularity. Noting that
 $0\leq a^2/\left(\beta+a^2\right)<1$ for $0\leq a<\infty$, we
 also require $\lambda<2$ to ensure $n^2>0$. Substituting
 Eq.~(\ref{eq33}) into Eq.~(\ref{eq10}), we find that the
 solution is given by
 \be{eq34}
 H=H_0\,a^{-3/2}\left[\frac{2\beta+\left(2-\lambda\right)a^2}{2\beta
 +2-\lambda}\right]^{\frac{\lambda}{4\left(2-\lambda\right)}}.
 \ee
 Obviously, $H\propto a^{-3/2}$ when $a\to 0$, and hence this model
 can describe the matter-dominated epoch. Then, we have
 \be{eq35}
 E^2=\left(1+z\right)^3\left[\frac{2\beta+\left(2-
 \lambda\right)\left(1+z\right)^{-2}}{2\beta+2
 -\lambda}\right]^{\frac{\lambda}{2\left(2-\lambda\right)}}.
 \ee
 On the other hand,
 \be{eq36}
 \Omega_m=\frac{\Omega_{m0}a^{-3}}{E^2}=\Omega_{m0}\left[
 \frac{2\beta+2-\lambda}{2\beta+\left(2-\lambda\right)
 \left(1+z\right)^{-2}}\right]^{\frac{\lambda}{2\left(2-
 \lambda\right)}}.
 \ee
 Again, requiring $\Omega_m\to 1.0$ (namely matter dominated)
 when $z\to\infty$, it is easy to find that
 \be{eq37}
 \Omega_{m0}=\left(\frac{2\beta}{2\beta+2-\lambda}
 \right)^{\frac{\lambda}{2\left(2-\lambda\right)}}.
 \ee
 Therefore, $\Omega_{m0}$ is {\em not} an independent parameter. And
 then, we have
 \be{eq38}
 \Omega_m=\left[\frac{2\beta}{2\beta+\left(2-\lambda\right)
 \left(1+z\right)^{-2}}\right]^{\frac{\lambda}{2\left(2-
 \lambda\right)}},
 \ee
 whereas $\Omega_\Lambda=1-\Omega_m$. Substituting Eq.~(\ref{eq34})
 into Eq.~(\ref{eq7}), we obtain the deceleration parameter
 \be{eq39}
 q\equiv -\frac{\ddot{a}}{aH^2}=\frac{1}{2}\left[
 1-\frac{\lambda a^2}{2\beta+\left(2-\lambda\right)a^2}\right].
 \ee
 It is easy to see that the universe is decelerating when
 $\left(\lambda-1\right)a^2<\beta$,
 whereas the universe is accelerating when
 $\left(\lambda-1\right)a^2>\beta$.
 The transition occurs when $\left(\lambda-1\right)a^2=\beta$.
 Therefore, $\lambda>1$ is necessary to accelerate the
 universe. The transition redshift is given by
 \be{eq40}
 z_t=\sqrt{\frac{\lambda-1}{\beta}}-1.
 \ee
 The effective EoS is given by (see e.g.~\cite{r21,r37})
 \be{eq41}
 w_{\rm eff}\equiv\frac{p_{tot}}{\rho_{tot}}
 =-1-\frac{2}{3}\frac{\dot{H}}{H^2}=\frac{1}{3}
 \left(2q-1\right)=-\frac{\lambda a^2}{\,3\left[2\beta
 +\left(2-\lambda\right)a^2\right]}\,,
 \ee
 whereas the EoS of HDE reads
 \be{eq42}
 w_\Lambda=\frac{w_{\rm eff}}{\Omega_\Lambda}
 =\frac{w_{\rm eff}}{1-\Omega_m}\,,
 \ee
 in which $\Omega_m$ is given in Eq.~(\ref{eq38}).

\begin{center}
 \begin{figure}[tbhp]
 \centering
 \includegraphics[width=0.55\textwidth]{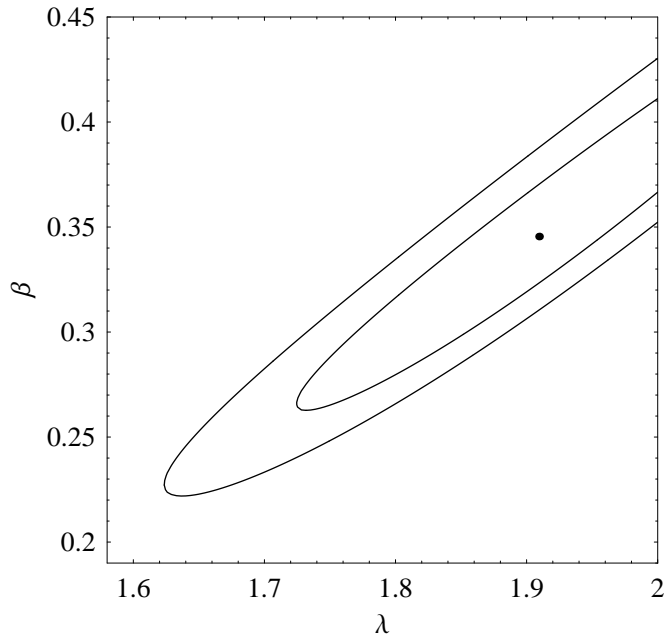}
 \caption{\label{fig3}
 The $68\%$ and $95\%$ confidence level contours in the
 $\lambda-\beta$ parameter space for the parameterization
 given in Eq.~(\ref{eq33}). The best-fit parameters are
 also indicated by a solid point.}
 \end{figure}
 \end{center}

\begin{center}
 \begin{figure}[tbhp]
 \centering
 \includegraphics[width=0.97\textwidth]{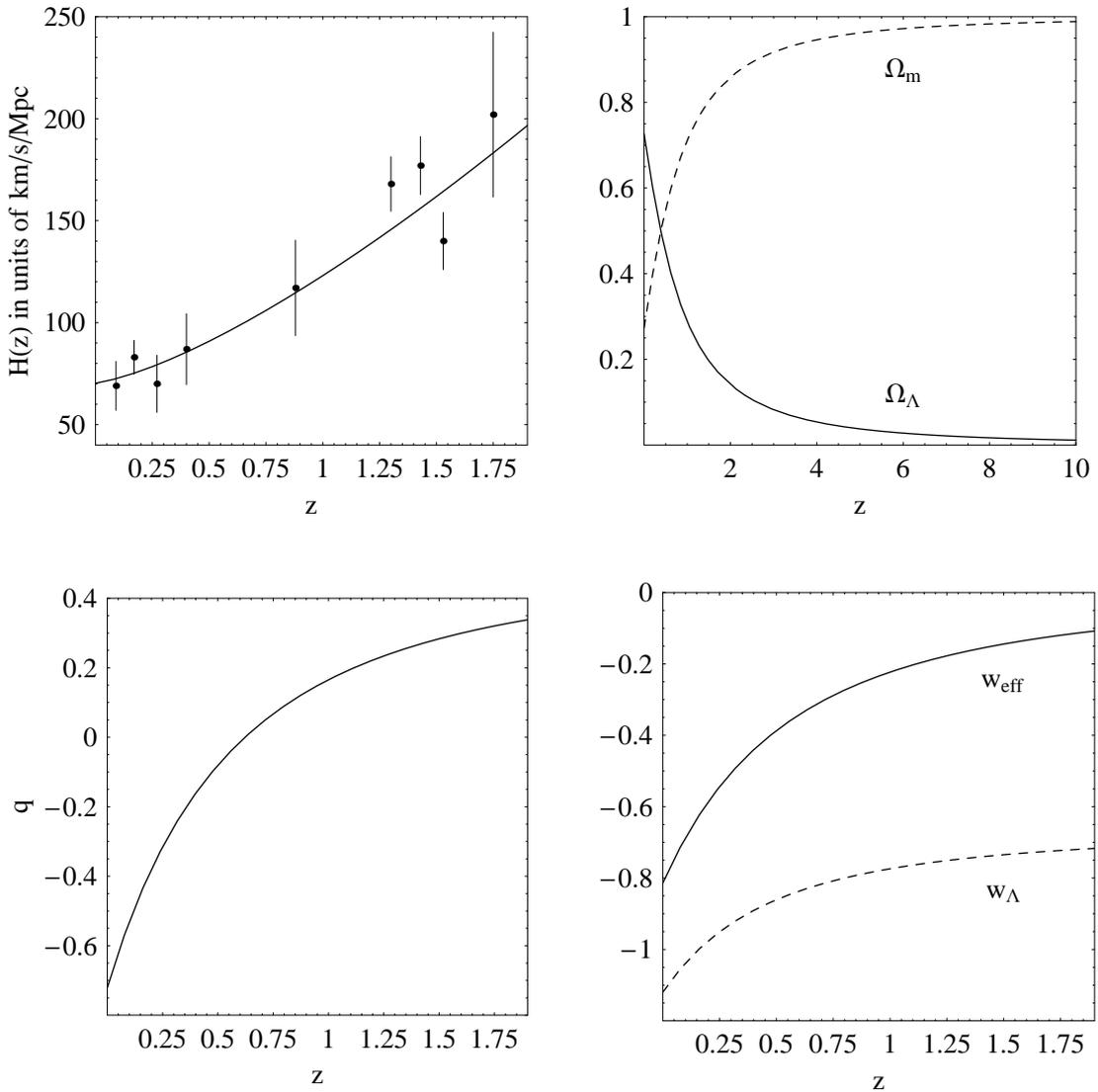}
 \caption{\label{fig4}
 The $H$, $\Omega_m$, $\Omega_\Lambda$, $q$, $w_{\rm eff}$ and
 $w_\Lambda$ as functions of redshift $z$ for the best-fit
 $\lambda$ and $\beta$ for the parameterization given in
 Eq.~(\ref{eq33}). These results are obtained by using the
 combined data of the 307 Union SNIa, the shift parameter $R$
 of CMB and the distance parameter $A$ of BAO.}
 \end{figure}
 \end{center}

Similar to Sec.~\ref{sec4}, we perform a joint $\chi^2$ analysis to
 obtain the cosmological constraints on the model parameters
 $\lambda$ and $\beta$, by using the combined data of the 307 Union
 SNIa, the shift parameter $R$ of CMB and the distance parameter $A$
 of BAO. By minimizing the total $\chi^2$ given in Eq.~(\ref{eq32}),
 we find that the best-fit parameters are $\lambda=1.9095$ and
 $\beta=0.3455$ (the corresponding $h=0.7024$), while
 $\chi^2_{min}=311.368$. In Fig.~\ref{fig3}, we present the
 $68\%$ and $95\%$ confidence level contours in the $\lambda-\beta$
 parameter space. We also plot $H$, $\Omega_m$, $\Omega_\Lambda$,
 $q$, $w_{\rm eff}$ and $w_\Lambda$ as functions of redshift $z$ for
 the best-fit $\lambda$ and $\beta$ in Fig.~\ref{fig4}. Obviously,
 the MHDE model with the parameterization given in Eq.~(\ref{eq33})
 is well consistent with the observational data in fact.


\vspace{-10mm}  

\begin{center}
 \begin{figure}[tbhp]
 \centering
 \includegraphics[width=0.869\textwidth]{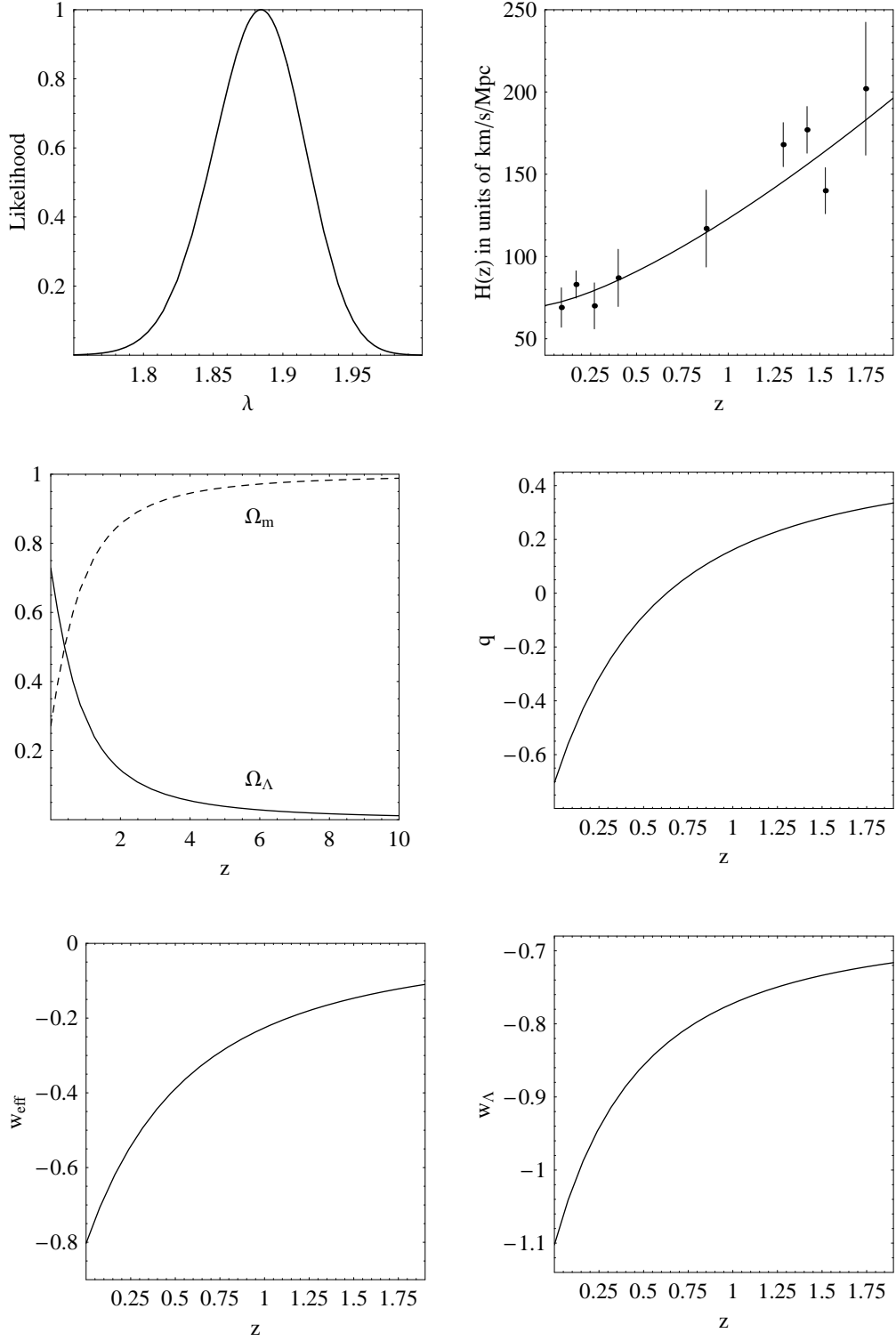}
 \caption{\label{fig5}
 The same as in Fig.~\ref{fig2}, except for the
 parameterization given in Eq.~(\ref{eq43}).}
 \end{figure}
 \end{center}

Note that there are two parameters in the parameterization
 given in Eq.~(\ref{eq33}). From the viewpoint of model
 building, it is better to find a single-parameter model.
 Inspired by the above results, we can consider another
 parameterization
 \be{eq43}
 n^2=2-\frac{3\lambda a^2}{1+3a^2}\,,
 \ee
 in which there is only one model parameter $\lambda$. In fact,
 it is just the reduced version of Eq.~(\ref{eq33}) with
 $\beta=1/3$. So, all the physical quantities of the MHDE model
 with this parameterization can be obtained by simply setting
 $\beta=1/3$ in Eqs.~(\ref{eq34})---(\ref{eq42}). Fitting to
 the combined data of the 307 Union SNIa, the shift parameter
 $R$ of CMB and the distance parameter $A$ of BAO, we find that
 the best-fit parameter reads $\lambda=1.8843^{+0.0318}_{-0.0336}$
 (with $1\sigma$ error) or $\lambda=1.8843^{+0.0619}_{-0.0692}$
 (with $2\sigma$ error), while $\chi^2_{min}=311.405$. The
 corresponding $h=0.7016$. In Fig.~\ref{fig5}, we present the
 corresponding likelihood ${\cal L}\propto e^{-\chi^2/2}$
 versus $\lambda$, as well as the $H$, $\Omega_m$,
 $\Omega_\Lambda$, $q$, $w_{\rm eff}$ and $w_\Lambda$ as
 functions of redshift $z$ for the best-fit $\lambda$.
 Obviously, the MHDE model with the parameterization
 given in Eq.~(\ref{eq43}) is also well consistent with the
 observational data.

Comparing to the parameterization (\ref{eq12}), besides
 the advantage of being valid for the whole $0\leq a<\infty$,
 the parameterizations (\ref{eq33}) and (\ref{eq43}) have
 smaller $\chi^2_{min}$ when we fit them to the combined data
 of the 307 Union SNIa, the shift parameter $R$ of CMB and the
 distance parameter $A$ of BAO.

\vspace{-10mm}  


\section{Conclusion and discussion}\label{sec6}
In this work, motivated by the energy bound suggested by Cohen
 {\it et al.}, we propose the modified holographic dark energy
 (MHDE) model. Choosing the IR cut-off $L=R_{\rm CC}$ and
 considering the parameterizations (\ref{eq12}), (\ref{eq33})
 and (\ref{eq43}), we derive all the physical quantities of the
 non-saturated MHDE model analytically. We find that the
 non-saturated MHDE models with parameterizations (\ref{eq12})
 and (\ref{eq43}) are single-parameter models in practice.
 Also, we consider the cosmological constraints on the
 non-saturated MHDE, and find that it is well consistent with
 the observational data in fact. Given the simplicity and
 analyticity, the non-saturated MHDE model is very interesting
 and deserves further investigation.

After all, there are some remarks on the parameterizations
 of $n(a)$. Firstly, our situation is similar to the familiar
 parameterization for the EoS of dark energy considered
 extensively in the literature, such as the well-known
 $w(a)=w_0+w_a (1-a)$~\cite{r20} and $w(z)=w_0+w_1 z$ (see
 e.g.~\cite{r39,r27,r40}). Until we can completely understand
 the mysterious dark energy in the far future, the method of
 parameterization is useful to shed some light on the nature
 of dark energy. Secondly, as mentioned in Sec.~\ref{sec3}, our
 parameterization $n^2=2-\lambda a$ given in Eq.~(\ref{eq12})
 is motivated by the EoS parameterization $w(a)=w_0+w_a (1-a)$.
 However, as recently argued by Shafieloo {\it et al.}~\cite{r41},
 the EoS parametrization $w(a)=w_0+w_a (1-a)$ cannot be trusted
 in the redshift interval $0<z<1.4$, and it fails for $a\gg 1$.
 Accordingly, one might doubt our parameterization
 $n^2=2-\lambda a$. As shown in Sec.~\ref{sec4}, the
 parameterization $n^2=2-\lambda a$ works well for $0<a\leq 1$,
 or, $0\leq z<\infty$. However, as mentioned in the beginning
 of Sec.~\ref{sec5}, it is only valid for $a<2/\lambda$ which
 is required by $n^2>0$. Thus, we proposed other
 parameterizations for $n^2$ in Eqs.~(\ref{eq33}) and (\ref{eq43})
 which work well for the whole $0\leq a<\infty$. Finally, one might
 argue that to be holographic (i.e., the energy density is
 proportional to the horizon area), $n(a)$ should be nearly
 constant. Therefore, $n(a)$ should not change so quickly. We
 admit that our parameterizations (\ref{eq12}), (\ref{eq33})
 and (\ref{eq43}) do not fulfill this point. On the other hand,
 they can be used as working parameterizations since they are
 well consistent with the observational data in fact. Nevertheless,
 it is still worthwhile to seek a more satisfactory parameterization
 for $n(a)$ which changes not so quickly and can be consistent with
 the observational data. We leave this to the future work. This
 tells us that there are still many improvements to do for the
 MHDE model.


\section*{ACKNOWLEDGEMENTS}
We thank the anonymous referee for quite useful comments and
 suggestions, which help us to improve this work. We are
 grateful to Professors Rong-Gen~Cai, Shuang-Nan~Zhang, and
 Miao~Li for helpful discussions. We also thank Minzi~Feng,
 as well as Chang-Jun~Gao, Xin~Zhang, Pu-Xun~Wu, Shi~Qi,
 Shuang~Wang and Xiao-Dong~Li, for kind help and discussions.
 This work was supported by the Excellent Young Scholars
 Research Fund of Beijing Institute of Technology.

\renewcommand{\baselinestretch}{1.41}


\end{document}